\documentclass{iopart}
\pdfoutput=1
\usepackage{iopams} 
\usepackage{mathptmx}
\usepackage[utf8]{inputenc} 
\usepackage{graphicx}
\usepackage{epstopdf}
\usepackage[numbers,comma,square,sort&compress,merge]{natbib} 
\usepackage{upgreek} 
\usepackage{xspace}

\usepackage{url}
\usepackage{color,soul} 

\usepackage[pdftex]{hyperref}
\hypersetup{
	pdftitle={Quenching of the luminescence intensity of GaN nanowires under electron beam exposure},
	colorlinks,
	citecolor=blue
	}

\newcommand{\celsius}{$^{\circ}$C\xspace}

\begin{document}
\graphicspath{{figures}}

\title[Quenching of the luminescence intensity of GaN nanowires under electron beam exposure]{Quenching of the luminescence intensity of GaN nanowires under electron beam exposure: Impact of C adsorption on the exciton lifetime}
\author{Jonas L\"ahnemann\footnote{Present address: Institut Nanosciences et Cryogénie (INAC-PHELIQS), CEA, 17 av.\ des Martyrs, 38000 Grenoble, France}, Timur Flissikowski, Martin W\"olz\footnote{Present address: Jenoptik Laser GmbH, Göschwitzer Str. 29, Jena, Germany}, Lutz Geelhaar, Holger T. Grahn, Oliver Brandt and Uwe Jahn}
\ead{laehnemann@pdi-berlin.de}
\address{Paul-Drude-Institut für Festkörperelektronik, Leibniz-Institut im Forschungsverbund Berlin e.\,V., Hausvogteiplatz 5--7, 10117 Berlin, Germany}

\begin{abstract}
Electron irradiation of GaN nanowires in a scanning electron microscope strongly reduces their luminous efficiency as shown by cathodoluminescence imaging and spectroscopy. We demonstrate that this luminescence quenching originates from a combination of charge trapping at already existing surface states and the formation of new surface states induced by the adsorption of C on the nanowire sidewalls. The interplay of these effects leads to a complex temporal evolution of the quenching, which strongly depends on the incident electron dose per area. Time-resolved photoluminescence measurements on electron-irradiated samples reveal that the carbonaceous adlayer affects both the nonradiative and the radiative recombination dynamics.  
\end{abstract}



\submitto{\NT}

\ioptwocol

\maketitle

\section{Introduction}

The large surface-to-volume ratio of semiconductor nanowires leads to a predominance of surface effects in their optical and electrical characteristics \cite{Calarco_nl_2005, vanWeert_apl_2006, Reshchikov_jvstb_2009, Pfuller_prb_2010,  Lahnemann_nl_2016, Calarco_jmr_2011, Grabowska_prb_2005}. In various material systems, both the conductivity and the emission properties have been shown to be governed by surface band bending induced by Fermi level pinning at the nanowire side walls \cite{Calarco_nl_2005, vanWeert_apl_2006, Reshchikov_jvstb_2009}. In unintentionally doped GaN nanowires with diameters below 100~nm, this band bending extends throughout the cross-section of the wire \cite{Calarco_nl_2005}. Another consequence of the large surface area is the observation of surface excitons, since the emission from excitons bound to donors located in the proximity of the surface is blue-shifted compared to that of bulk-like donor-bound excitons \cite{Grabowska_prb_2005,Brandt_prb_2010, Corfdir_jap_2009b}. Furthermore, spectroscopic measurements themselves may (unintentionally) affect the electronic properties of the wires \cite{Pfuller_prb_2010,Wallys_nl_2012}. For example, an extended exposure of GaN nanowires placed in vacuum or in a N$_2$ atmosphere to ultraviolet radiation leads to a significant increase in the photoluminescence (PL) intensity, which is due to a photoinduced desorption of O$_2$. This process results in an unpinning of the Fermi level and a corresponding reduction of the radial electric fields associated with surface band bending \cite{Pfuller_prb_2010}.

An effect to the contrary is observed when GaN nanowires are investigated by cathodoluminescence (CL) spectroscopy necessitating their exposure to an electron beam. In these experiments, a strong decrease of the near band-edge emission is inevitably observed. This quenching has been reported in the literature \cite{Robins_jap_2007a, Lim_nl_2009}, but no conclusive explanation has been given. Robins \etal~\cite{Robins_jap_2007a} have considered charge trapping effects, C contamination, and the formation of point defects as possible origins of the quenching. They argued that neither of the latter two possibilities are likely to be the reason for the quenching, and solely attributed the quenching to charge trapping, which they proposed to enhance the nonradiative recombination rate. Similarly, for unintentionally doped GaN layers, a (less significant) quenching of the CL signal with time has been reported \cite{Campo_mrssp_2001, Campo_ijnsr_2004, Wang_jvsta_2009}. Absorption of the emitted CL in an electron beam deposited carbonaceous layer as a possible explanation was discarded on the basis of energy- and wavelength-dispersive x-ray spectroscopy \cite{Campo_ijnsr_2004}. However, these techniques are not sensitive enough to detect very thin carbonaceous layers, whereas already a sub-monolayer C coverage could alter surface states. In fact, Wang \etal ~\cite{Wang_jvsta_2009} proposed that adsorbed species on the surface play a role in the CL quenching in GaN layers, without further elucidating the nature of these adsorbates. A quenching of the CL intensity in nanostructures with progressive electron beam exposure has been reported for other semiconductors as well, particularly for ZnO \cite{Dierre_jap_2008b}.

We have systematically observed this quenching process during CL measurements on a variety of GaN nanowire samples grown in different molecular beam epitaxy reactors both in our laboratory and by external collaborators. The quenching can be observed both at room temperature and at cryogenic sample temperatures and, though its magnitude varies from sample to sample, amounts to a reduction in emission intensity by one to two orders of magnitude. In the present work, we systematically analyze the quenching behavior observed in CL measurements at room temperature, thereby relating it to the nanowire surface with contributions from both charging and C deposition. C contamination of the sample surfaces in a scanning electron microscope (SEM) is a long-known problem \cite{Ennos_bjap_1953,Hart_pm_1970}. Modern instruments employ oil-free vacuum pumps to reduce the abundance of hydrocarbons in the SEM chamber, particularly those of higher molecular weight. However, unless an ultra-high vacuum system is used, the partial pressure of hydrocarbons in typical SEM chambers is still considerable. The electron beam will crack hydrocarbons sticking to the specimen surface, resulting in the gradual deposition of a thin carbonaceous film. Polar molecules may actually be attracted to the exposed area as a result of electric fields induced by charging from the electron irradiation and enhance the C deposition \cite{Hart_pm_1970,Fourie_optik_1978}. Thus, the effects of surface charging and C deposition are interlinked.

\section{Experimental details}

The nominally undoped, self-assembled GaN nanowire ensembles investigated in this study were grown on Si(111) substrates by plasma-assisted molecular beam epitaxy (MBE) under N-rich conditions (N/Ga flux ratio of 5) at a substrate temperature of 780~\celsius. The total growth time was 90~min, yielding nanowires with an average length of 530~nm and an average equivalent disk diameter \cite{Brandt_cgd_2014} of $\left<d\right>=\left(50 \pm 20\right)$~nm at a density of about $1\times 10^{10}$~cm$^{-2}$. To investigate the role of the surface for the luminescence quenching, the nanowire growth was repeated with nominally identical conditions, after which the GaN core was covered by an (Al,Ga)N shell for an additional 5~min. To promote radial growth, we kept the temperature constant but increased the total group-III metal flux such that flux ratios of $\mathrm{N/III}=1.15$ and $\mathrm{Al/Ga}=0.4$ were obtained. The average diameter of the core/shell nanowire tips was found to be 10~nm larger than that of the sample without shell, indicating a shell thickness of about 5~nm at the top of the nanowires, which is reduced toward the base of the nanowires due to the shadowing of the incoming molecular beam. The Al$_{x}$Ga$_{1-x}$N composition was estimated as $x = 0.2\pm 0.1$ from a Gaussian fit to the (Al,Ga)N peak in CL spectra of the nanowire ensemble, which is slightly lower than the nominal value of $x \approx 0.29$ determined from the metal flux ratio during growth.


CL measurements were carried out with a Gatan Mo\-noCL3 system fitted to a Zeiss Ultra55 field-emission SEM \cite{Lahnemann_jpd_2014}. For simplicity, we limit this investigation to CL measurements at room temperature. The SEM was operated at 5 or 10~kV with beam currents between 0.5 and 6.5~nA. The nanowire luminescence was collected using a parabolic mirror and dispersed using a spectrometer (focal length of 30~cm, 300 or 1200~lines/mm gratings, 0.1--0.5~mm slits). The monochromatized light was detected using a photomultiplier tube (PMT). The spectra were recorded using a charge-coupled device (CCD) as the detector. For measurements with extended electron beam exposure, the microscope was set to scan a certain (freshly-exposed) field of view at the fastest scan speed (122.3 ms/frame), while the monochromatic CL intensity was continuously recorded with the PMT (integrated over intervals of 0.5 or 1~s). 

The effect of an ex-situ 2~min H$_2$/O$_2$ plasma cleaning step preceding the insertion of the samples into the SEM was investigated using a Gatan Solarus (model 950) system at a power of 50~W with H$_2$ and O$_2$ fluxes of 6.4~sccm and 27.5~sccm, respectively. Intentional sputter deposition of C was carried out in a Gatan precision etching and coating system.

Continuous-wave PL spectra were recorded at room temperature using a confocal $\upmu$-PL setup with excitation from the 325-nm line of a He-Cd laser and a CCD for detection. Time-resolved $\upmu$-PL experiments were performed at room temperature by focusing the second harmonic (325~nm) of an optical parametric oscillator pumped by a femtosecond Ti:sapphire laser onto the sample. The PL signal was dispersed by a monochromator and detected by a streak camera operating in synchroscan mode. 

\begin{figure}
\centering
\includegraphics*[width=7.5cm]{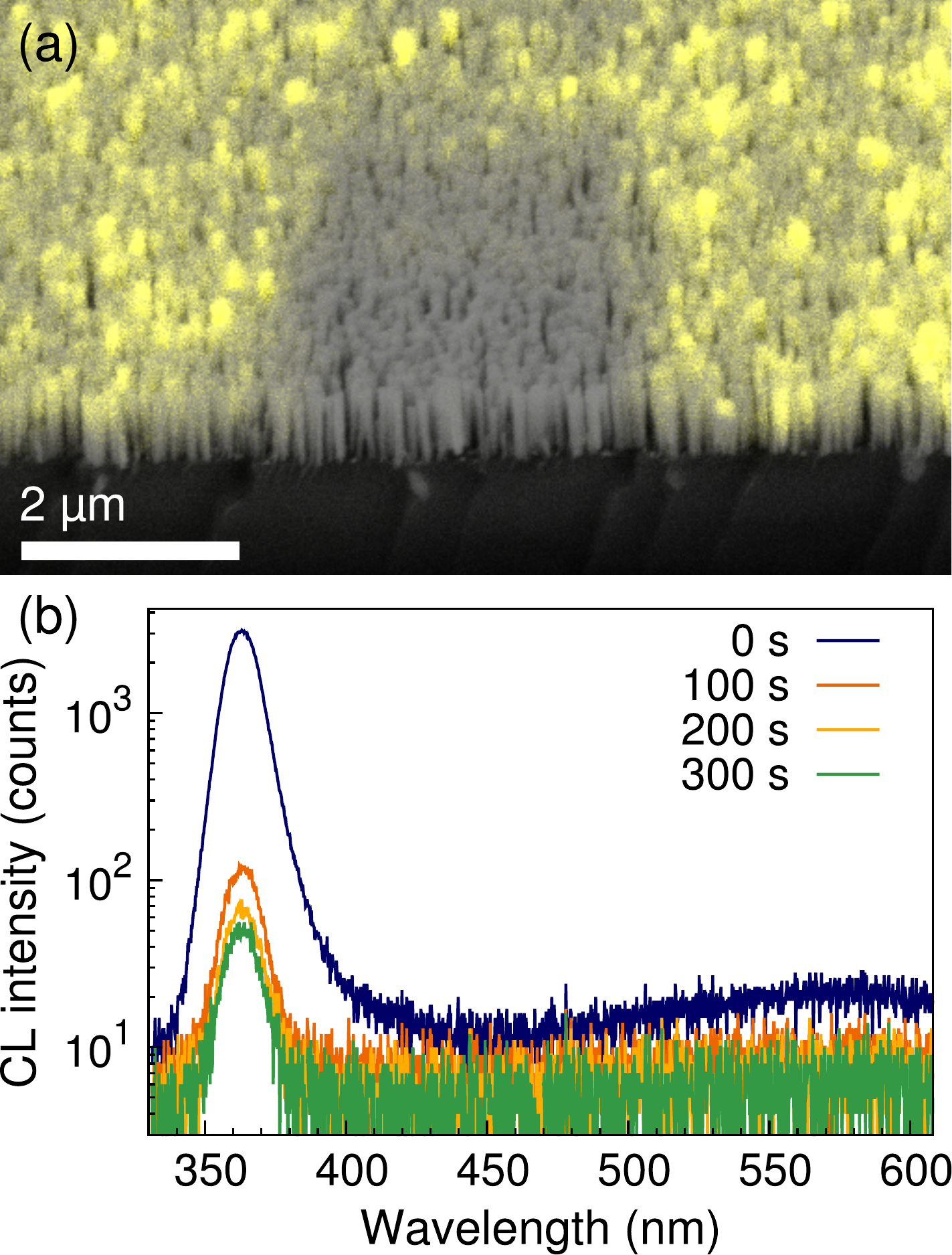}
\caption{\label{fig:illustr}(a) Superposition of a panchromatic CL image of the GaN emission intensity (yellow) at room temperature with the corresponding bird's eye view SEM image of the nanowires. The central part has been exposed to the electron beam for several tens of seconds before zooming out to record the images (beam current 0.1~nA). (b) Room temperature CL spectra recorded in top-view geometry upon first exposure ($t=0$~s) and on the same area after 100, 200, and 300~s of electron beam exposure (scanned area 98~$\upmu$m$^2$, beam current 0.75~nA).}
\end{figure}

\section{Results and discussion}

An illustration of the CL quenching is given in figure~\ref{fig:illustr}(a), where a bird's eye view CL image was recorded on the cross-section of a GaN nanowire ensemble directly after reducing the magnification so that the center of the image was exposed to the electron beam for several tens of seconds, while the rest of the image was freshly exposed. The freshly exposed part shows strong near band-edge (NBE) emission due to the radiative decay of free excitons, while the other part was pre-exposed to the electron beam and, in comparison, appears basically dark. Room temperature CL spectra recorded in top view on the nanowire ensemble with a CCD detector for a freshly-exposed field (0~s) and after scanning the same area for up to 300~s are shown in figure~\ref{fig:illustr}(b). The intensity of the NBE emission is reduced without apparent spectral shift, change in peak shape, or redistribution of intensity to deep level defect transitions.

\begin{figure}
\centering
\includegraphics*[width=7.5cm]{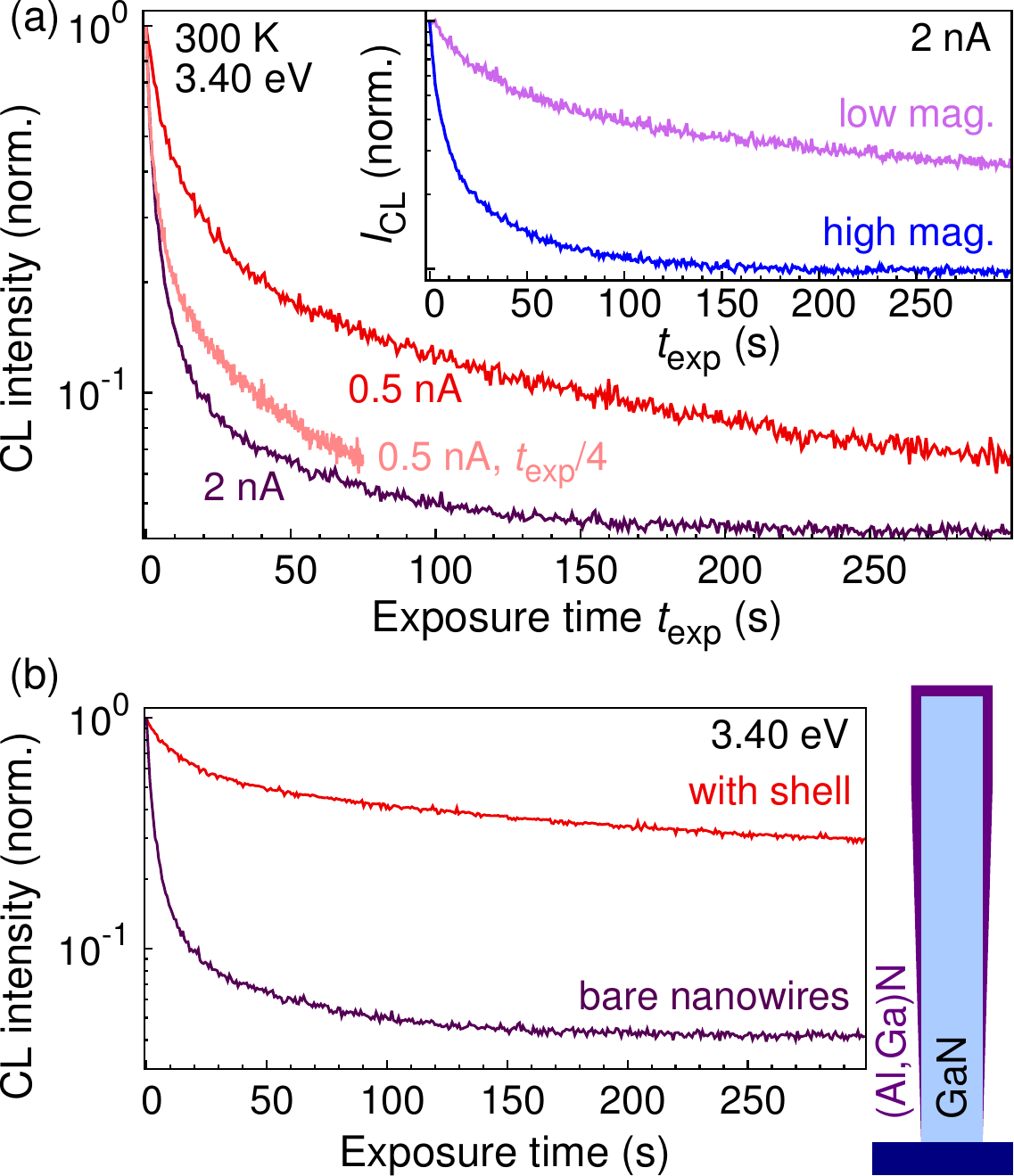}
\caption{\label{fig:surface}(a) Temporal evolution of the normalized CL intensity $I_{CL}$ at room temperature (note the logarithmic intensity scale) under electron beam irradiation for a GaN nanowire ensemble continuously scanned in top-view geometry (scanned area 98~$\upmu$m$^2$) using two different beam currents. For comparison, the transient for the lower beam current (0.5 nA) is also plotted with the time scaled by a factor 1/4. The inset shows two CL transients measured at low (scanned area 2450~$\upmu$m$^2$) as well as high (scanned area 98~$\upmu$m$^2$) magnification. (b) CL quenching of the bare nanowire ensemble compared with that for nanowires covered by an (Al,Ga)N shell (scanned area 98~$\upmu$m$^2$, beam current 2 nA). The sketch to the right schematically depicts the geometry of the core/shell nanowire heterostructure.}
\end{figure}

A quantitative analysis of the luminescence quenching is made possible by recording the monochromatic CL intensity over time while continuously scanning a fixed sample area (in top-view geometry). Unless otherwise noted, the luminescence intensity refers to the spectral maximum of the NBE emission of GaN (364~nm or 3.4~eV). Figure~\ref{fig:surface} shows several such measurements carried out under different exposure conditions, where the intensity after 300~s of exposure can decrease by as much as a factor of 25. The measurements in figure~\ref{fig:surface} provide us with several insights into the quenching process. 

First, we show measurements at different beam currents (0.5 and 2~nA) in figure~\ref{fig:surface}(a) and at different magnifications [inset of figure~\ref{fig:surface}(a)], where an increase in the magnification also increases the electron dose per area. In both cases, an increase of the incident electron dose accelerates the quenching. In fact, scaling the time axis by the ratio in beam currents [figure~\ref{fig:surface}(a)] yields quite similar curves and shows that the process roughly scales with the electron dose. Therefore, the quenching is a process depending on the incident electron dose. At a high impinging electron dose, the transients are characterized by a strong initial drop during the first 30~s, which subsequently slows down significantly. At a reduced electron dose, both the initial drop and the slow-down are not as pronounced. Note that the time dependence of the quenching process is non-exponential, but neither follows a stretched exponential nor a power law. 

Second, to demonstrate the role of the sidewall surfaces, we have investigated the quenching in an ensemble of GaN nanowires covered with a thin (Al,Ga)N shell. This wide gap passivation layer prevents carriers within the GaN core from reaching the surface. In figure~\ref{fig:surface}(b), the luminescence quenching in the bare nanowire ensemble is compared to the nanowires covered by an (Al,Ga)N shell. The reduction of the luminescence intensity is much less significant for the latter sample: over a time period of 300~s, the intensity is reduced by a factor of 25 for the bare GaN nanowires, but only by a factor of 3 for the core/shell nanowires. This measurement unambiguously confirms that we are dealing with a surface-related effect. Note that the importance of the surface for the quenching process is already inferred by the fact that the quenching reported for layers under similar conditions is not as significant \cite{Campo_mrssp_2001, Wang_jvsta_2009}.

In order to further elucidate the origin of the observed electron-beam-induced quenching of the luminescence intensity without any alteration of the spectral shape, a number of possibilities have to be considered \cite{Robins_jap_2007a}: the generation of point defects by radiation damage, the trapping of injected charges at surface states, and the adsorption of C on the nanowire surface.

\subsection{Radiation damage}

In principle, the quenching could originate from the generation of native point defects acting as nonradiative recombination centers. However, to generate point defects in the N and Ga sublattices of a GaN crystal, one typically has to employ electron irradiation at 450 and 2000~keV, respectively, with the threshold values for defect formation lying at about one third of these energies \cite{Tuomisto_prb_2007}. At the lower acceleration voltages typical for SEM imaging (i.\,e., 5--10~keV), the activation of H-passivated Mg acceptors \cite{Amano_jjap_1989} or intrinsic point defects \cite{Nykanen_apl_2012} has been reported to affect the luminescence intensity. However, in contrast to samples grown by metal-organic vapor phase epitaxy or reactive MBE employing NH$_3$ as the N precursor, H is not incorporated in significant quantities during plasma-assisted MBE, the technique employed for the growth of the samples investigated in this study. In any case, the creation of point defects and the resulting non-radiative recombination would not be inhibited by the presence of an (Al,Ga)N shell, contrary to the observation in figure~\ref{fig:surface}(b). Furthermore, if the luminescence quenching were caused by the generation or activation of point defects within the nanowires, we would expect planar GaN layers to be affected to a similar degree as the nanostructures. Hence, we believe that radiation damage can be safely excluded as the origin of the luminescence quenching in GaN nanowire ensembles.

\begin{figure}
\centering
\includegraphics*[width=\columnwidth]{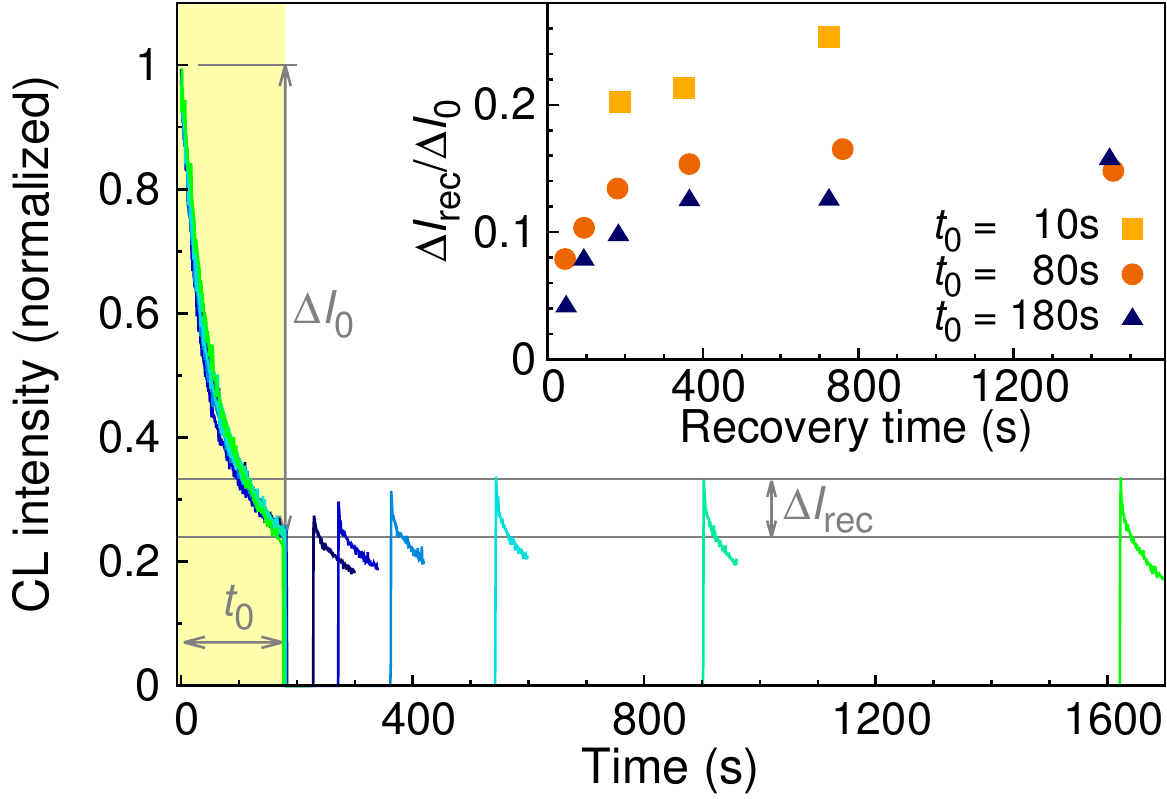}
\caption{\label{fig:recovery}Recovery of the quenched CL intensity with time during beam blanking. After a primary exposure time of 180~s (shaded yellow), the beam was blanked for 45~s to 24~min. The evolution of the signal upon re-exposure is plotted in different colors (scanned area: 1090~$\upmu$m$^2$, beam current: 5~nA). Each measurement was carried out on a previously unexposed area. In the inset, the recovery in CL intensity during the blanking period normalized to the intensity decay prior to blanking ($\Delta I_\mathrm{rec}/\Delta I_0$) is shown for three series of measurements with different primary exposure times $t_0=$ 10, 80, and 180~s.}
\end{figure}

\subsection{Charge trapping}

Next, we turn to the charging of surface states, a process which is expected to be largely reversible. For the oxidized \emph{M}-plane surface of GaN, the Fermi level is known to be pinned at about 0.6~eV below the conduction band edge \cite{Calarco_nl_2005}. Pf\"uller \etal~\cite{Pfuller_prb_2010} reported an unpinning of the Fermi level in GaN nanowires due to O$_2$ desorption from their sidewall surfaces under irradiation by an ultraviolet laser. This phenomenon results in a reduction of the radial electric fields, which in turn leads to an enhanced radiative decay rate (i.\,e., higher PL intensity) in the unpinned state. The charging of surface states may essentially have the opposite effect: changing the charge state of the surface states may pin the Fermi level closer to mid-gap, thus increasing the radial electric fields and reducing the radiative decay rate.

Robins \etal~\cite{Robins_jap_2007a} considered this charge trapping as the dominant source of CL quenching as they observe that a temperature cycling subsequent to a quenching measurement at low temperature leads to a recovery of the original CL intensity. They attributed this recovery to a thermally activated release of the trapped charges. However, in our own experiments, we observed only a partial recovery of the CL intensity after heating up a sample exposed to the electron beam at low temperature (10~K). In CL images recorded at room temperature, the contrast from the previously exposed region is still clearly darker than from its vicinity. Note in this context that, when remeasuring quenching transients on a previously exposed field as presented in  \cite{Robins_jap_2007a}, already a minor misalignment of the fields may lead to an overestimation of the recovery. 

To further investigate the non-persistent trapping of charges, we have investigated the recovery of the CL quenching during ``blanking'' of the electron beam (deflection from its usual path in the electron column) for intermittent periods while recording CL transients at room temperature. An example of such measurements is given in figure~\ref{fig:recovery}. Different fields were exposed to the electron beam for 180~s prior to blanking the beam. For an increasing blanking time, the signal intensity upon re-exposure increases. This recovery saturates after a few ten minutes of blanking. To quantify this recovery, we can define the relative recovery as $\Delta I_\mathrm{rec}/\Delta I_0$, where $\Delta I_\mathrm{rec}$ is the absolute recovery in CL intensity during the blanking period, which is normalized to the quenching of the CL intensity $\Delta I_0$ during the primary exposure (as illustrated by the arrows in figure~\ref{fig:recovery}). The relative recovery in CL intensity during beam blanking for several series of such measurements with different primary exposure times is summarized in the inset of figure~\ref{fig:recovery}.  The relative recovery is more pronounced for a short primary exposure (10~s), where it amounts to about 25\% after 10~min, than for longer primary exposures (80 and 180~s), where it reaches only about 16\% even after more than 20~min. This behavior indicates that the charge trapping occurs on shorter time scales than the persistent quenching, the nature of which will be investigated in the next section. Concerning the dynamics of the charge trapping, the gradual occupation of available surface states should continuously reduce the capture rate for electrons, since like charges repel. As a consequence, the process is expected to slow down with time, contributing to the complex non-exponential dynamics of the quenching process. Note that an incident electron from the beam creates about $10^3$ low-energy secondary electrons \cite{Mitsui_jjap_2005}. Hence, charge trapping can indeed be expected to be a rather fast process.

The partial reversibility of the quenching process confirms that charge trapping does contribute to the CL quenching. However, considering that the magnitude of the recovery does not exceed 25\%, it is clear that charge trapping cannot be the sole origin of the quenching, but is indeed of rather minor importance, in particular for longer exposure times.

\begin{figure}
\centering
\includegraphics*[width=7.5cm]{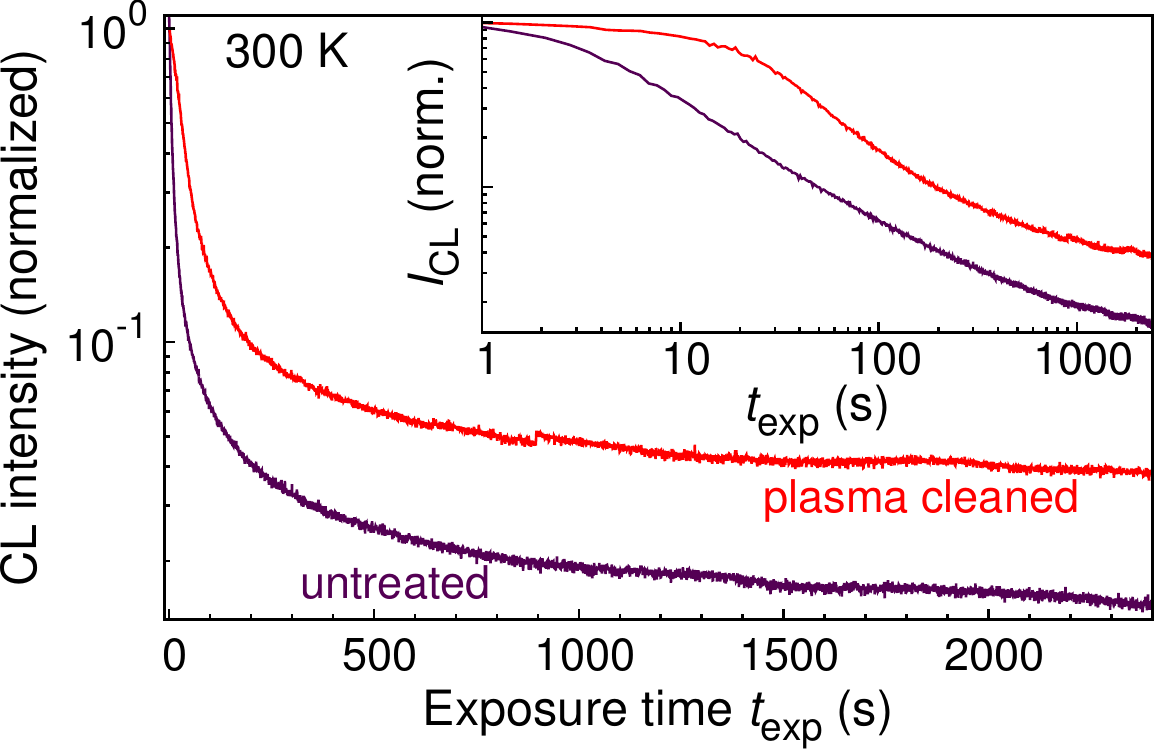}
\caption{\label{fig:plasma} Comparison of the evolution of the GaN NBE CL intensity with time for a sample piece treated in H$_2$/O$_2$ plasma for 2 min prior to the placement in the SEM and an untreated piece (scanned area 98~$\upmu$m$^2$, beam current 1.2~nA). The inset shows a double-logarithmic plot of the same data.}
\end{figure}

\subsection{Carbon adsorption}

A persistent quenching of the luminescence may result from the chemisorption of adsorbates. An obvious candidate in SEM-based measurements is C, which is deposited on investigated samples due to the cracking of hydrocarbon molecules present on the sample surface or in the vacuum chamber by the high-energy electron beam \cite{Ennos_bjap_1953,Hart_pm_1970,Fourie_optik_1978}. On the one hand, the build-up of a carbonaceous layer on the surface might simply lead to the absorption of the emitted luminescence in this opaque adlayer. On the other hand, the adsorption of C on the surface may induce surface states promoting nonradiative recombination. In contrast to the optical absorption, a sub-monolayer coverage could already be sufficient for giving rise to a pronounced quenching of the luminescence.

A pre-treatment of samples in a combined H$_2$/O$_2$ plasma is known to be an effective way to remove hydrocarbon contamination on the surface and thereby mitigate the formation of a carbonaceous layer during electron beam exposure \cite{Griffiths_jpcs_2010}. Figure \ref{fig:plasma} shows a comparison between the CL quenching with exposure time for an untreated and a plasma-cleaned piece of the same sample. Clearly, the quenching is less severe for the piece subjected to the 2~min plasma cleaning step. Particularly, the initial quenching is much slower as can be clearly seen in the double logarithmic plot in the inset of figure~\ref{fig:plasma}. For short exposure times, the remaining slow quenching for the cleaned sample piece is probably due to charging effects alone. However, the quenching observed for longer exposure times is still considerable, and the improvement gained by the plasma cleaning step amounts to only a factor of three. We believe this finding to be due to the abundance of hydrocarbons in the high vacuum of a typical SEM. A partial pressure of $10^{-8}$~Torr, for example, results in monolayer coverage within a few minutes. This view is confirmed by the fact that plasma treatments for longer than two minutes did not result in any further improvement. 

\begin{figure}
\centering
\includegraphics*[width=7.5cm]{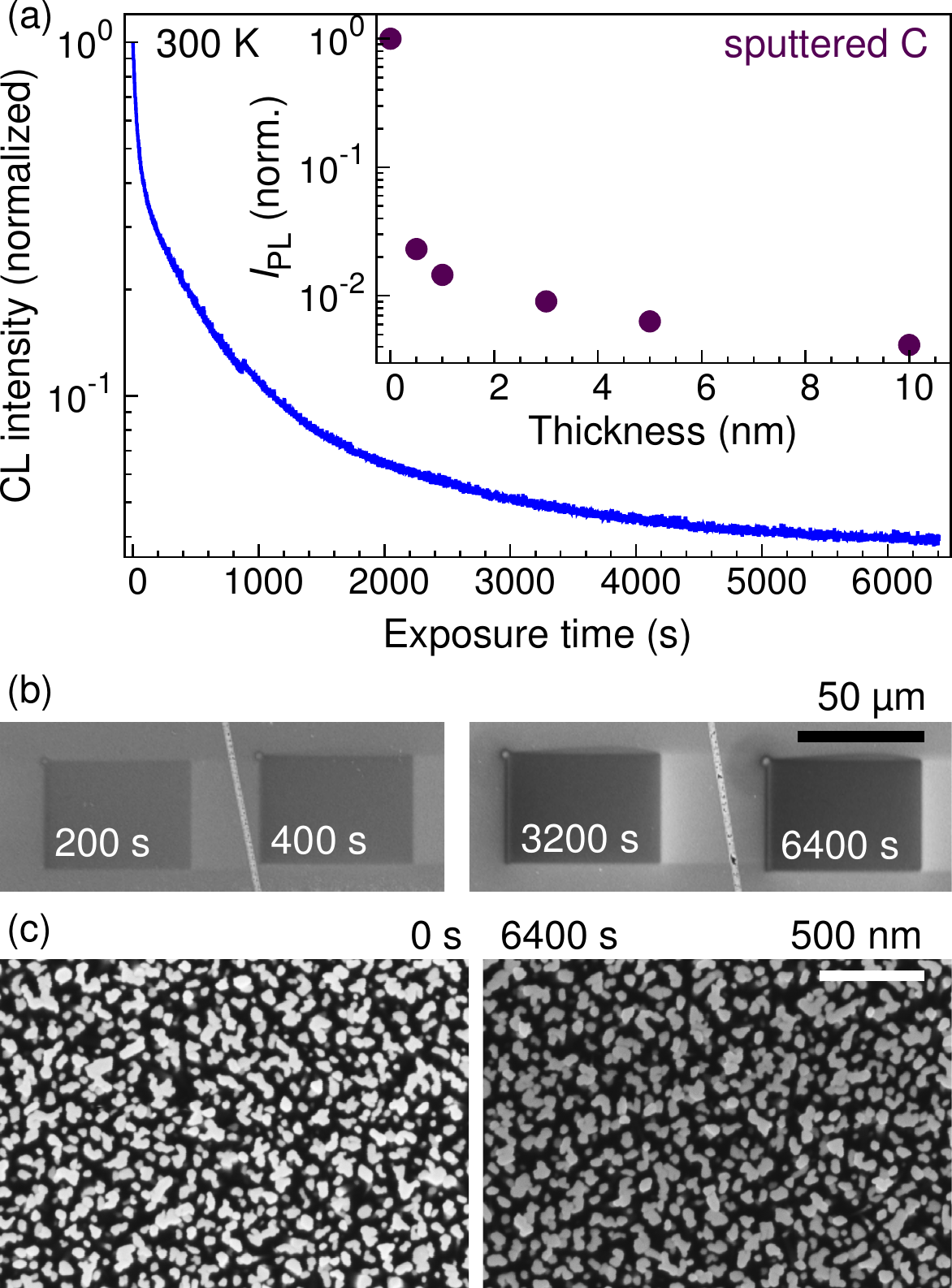}
\caption{\label{fig:exposure}(a) Evolution of the normalized GaN NBE CL intensity during a 6400-s exposure on a large field.
Inset: Normalized integrated PL intensity at room temperature as a function of nominal layer thickness for nanowire samples sputter coated with C. (b) SEM images (in-lens detector, same contrast settings) showing the material contrast from a carbonaceous film for four fields exposed to the electron beam in the SEM for different times. For each field, an area of 2450~$\upmu$m$^2$ was continuously scanned with a beam current of 6.5~nA. (c) Magnified top-view SEM images of unexposed nanowires (left) and nanowires from the field exposed for 6400~s (right).}
\end{figure}

To further elucidate the role of C adsorption in the process of luminescence quenching, we have carried out a series of electron beam exposures under controlled conditions. To this end, we have scanned large fields (2450~$\upmu$m$^2$) in the SEM at a fixed beam current of 6.5~nA and varied the exposure time from 200 to 6400~s. The CL quenching for the longest of these exposures is exemplified in figure~\ref{fig:exposure}(a), where shorter exposure times simply interrupt this curve at an earlier time. On these large fields, the quenching continues even after almost two hours of exposure since the dose is comparatively small. Figure~\ref{fig:exposure}(b) shows SEM images for four of the exposed fields recorded with the in-lens secondary electron detector. Such an in-lens detector is most sensitive to low-energy secondary electrons. The scattering of these low-energy electrons yields a strong chemical contrast enabling the detection of monolayer graphene films \cite{Kochat_jap_2011,Wofford_njp_2014}. The dark contrast visible in the images in figure~\ref{fig:exposure}(b) is thus consistent with the formation of a carbonaceous film.

The strong contrast observed after an exposure of 6400~s may lead to the impression that we have deposited a layer of significant thickness on the surface. In this case, an obvious candidate for the persistent quenching would be optical absorption. For amorphous C, the absorption coefficient at 3.4~eV has been measured to be around $2\times10^5$~cm$^{-1}$ \cite{Hagemann_jopt_1975,Mominuzzaman_jjap_1999}. A 10~nm thick C film would therefore absorb only about 20\% of the NBE CL from the nanowires. A similar absorption would be reached in the case of short-range ordered C layers for a thickness of 5~nm \cite{Dovbeshko_nrl_2015}. However, when examining the exposed nanowire ensemble at a higher magnification as shown in figure~\ref{fig:exposure}(c), no apparent change of the nanowire diameter is observed even for the longest exposure time. The area coverage of the nanowires in the two micrographs in figure~\ref{fig:exposure}(c) increases slightly from 49 to 51\%, which for circular nanowires with an average diameter of 50~nm would correspond to a diameter increase of 1~nm. However, the uncertainty in this measurement is actually larger than this difference.

Additionally, we have investigated sputter-coated nanowire samples with nominally 0.5--10~nm of C (corresponding to the layer measured by a quartz thickness monitor). The evolution of the room temperature PL intensity with nominal C thickness for these samples is given in the inset to figure~\ref{fig:exposure}(a). Already an equivalent C thickness of 0.5~nm is sufficient to quench the PL emission by almost two orders of magnitude, comparable to the quenching in CL measurements. Considering the much larger surface area of the nanowires, this thickness translates into a sub-monolayer coverage of their sidewalls. In consequence, it can be ruled out that optical absorption plays a significant role in the CL quenching. Instead, the quenching must be related to a change in the surface states induced by the presence of C. Furthermore, this measurement constitutes an independent proof that we can rule out the introduction of point defects and that also the charging of surface states plays only a subordinate role in the luminescence quenching.

\begin{figure}[t]
\centering
\includegraphics*[width=7.6cm]{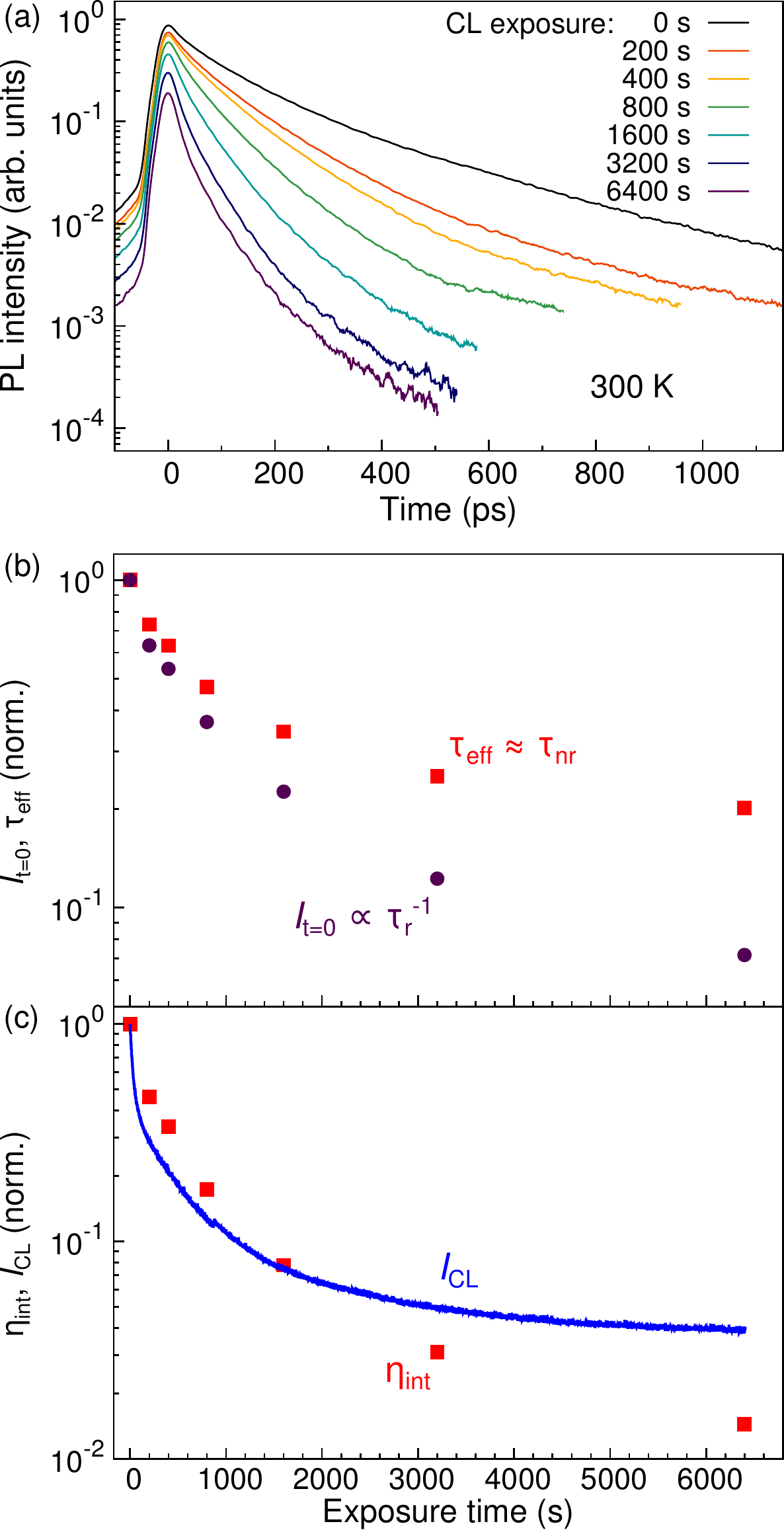}
\caption{\label{fig:trpl}(a) Room temperature PL transients taken from an unexposed area of the GaN nanowire ensemble (exposure 0~s) and from the six fields previously scanned in the SEM/CL for exposure times between 200 and 6400~s [\emph{cf} figure~\ref{fig:exposure}(b)]. The $1/e$ time $\tau_\mathrm{eff}$ of the initial decay decreases from 120~ps for the pristine nanowires to 25~ps for an exposure of 6400~s. (b) Evolution of the normalized effective lifetime  $\tau_\mathrm{eff} \approx \tau_\mathrm{nr}$ (squares) with exposure time compared with the corresponding normalized, spectrally integrated PL intensity at zero delay $I_\mathrm{t=0} \propto \tau_\mathrm{r}^{-1}$ (circles). (c) Normalized internal quantum efficiency $\eta_\mathrm{int}$ as determined from the time-resolved PL measurements in comparison to the CL intensity $I_\mathrm{CL}$ during the corresponding beam exposure.}
\end{figure}

This presence of C on the nanowire sidewall surfaces could, on the one hand, introduce nonradiative recombination centers at the surface, increasing the nonradiative decay rate. On the other hand, C adatoms may induce the formation of surface states that pin the Fermi level closer to midgap, thus further increasing the radial electric fields and thereby reducing the radiative recombination efficiency in analogy to the charge trapping at already existing surface states. For a direct quantitative determination of the radiative and nonradiative recombination rates, we have investigated the charge carrier dynamics by time-resolved micro-PL measurements on the fields depicted in figure~\ref{fig:exposure}(b). For these ex-situ PL measurements, we can expect that reversible charging of surface states does not play a role and that we are dealing only with persistent modifications of the surface states. 

The recorded PL transients are shown in figure~\ref{fig:trpl}(a). The effective exciton lifetime $\tau_\mathrm{eff}$, which we define as the 1$/e$ time of the initial decay, amounts to 120~ps for the pristine nanowire ensemble. This short lifetime is essentially purely nonradiative, but is not related to surface recombination since we have measured values exceeding 1~ns for other GaN nanowires of similar diameter. For extended exposures, the PL decay accelerates, and $\tau_\mathrm{eff}$ is eventually reduced to 25~ps for the maximum exposure time. This monotonic reduction of $\tau_\mathrm{eff}$ directly reflects a corresponding increase of the nonradiative recombination rate due to the electron-beam induced C deposition on the surface. Its clear correlation with the exposure time [\emph{cf} figure~\ref{fig:trpl}(b)] demonstrates that the adsorbed C atoms act as nonradiative recombination centers and drastically increase the surface recombination velocity. Since $\tau_\mathrm{eff} \approx \tau_\mathrm{nr} \ll \tau_\mathrm{r}$ with $\tau_\mathrm{r}$ denoting the radiative and $\tau_\mathrm{nr}$ the effective nonradiative lifetime, we can write $\tau_\mathrm{eff} \approx \tau_\mathrm{nr} = \left(1/\tau_\mathrm{nr}^b + 1/\tau_\mathrm{nr}^s\right)^{-1}$ with the nonradiative contributions from the bulk ($b$) and the surface ($s$). With $\tau_\mathrm{eff} = 25$~ps and $\tau_\mathrm{nr}^b = 120$~ps, we obtain $\tau_\mathrm{nr}^s = 32$~ps. For the present average nanowire diameter of  $\left<d\right> = 50$~nm, we can estimate the surface recombination velocity $S$ of the \emph{M}-plane sidewall surface of the nanowires after 6400~s of exposure to $S = 4 \left<d\right> / \tau_\mathrm{nr}^s = 6.2 \times 10^5$~cm/s. This value has to be compared to the one reported by Schlager \etal~\cite{Schlager_jap_2008} for as-grown GaN microwires at room temperature, namely, $S = 9\times10^3$~cm/s. Apparently, the deposition of C increases $S$ by almost two orders of magnitude, reaching a magnitude commonly associated with other semiconductors such as GaAs \cite{Joyce_nt_2013}.

Together with this decrease in $\tau_\mathrm{eff}$, the spectrally integrated peak intensity at zero delay ($I_{t=0}$) also decreases with the exposure time as clearly seen in figure~\ref{fig:trpl}(a). Since we have ruled out a significant absorption in the deposited C layer, this reduction of $I_{t=0}$ implies that the electron beam exposure affects not only the nonradiative, but also the radiative decay rate. For excitation pulses much shorter than the recombination time, $I_{t=0}$ is directly proportional to the radiative decay rate: $I_{t=0} \propto \tau_\mathrm{r}^{-1}$ (see  \cite{Brandt_jvstb_2002} for an explicit derivation). The evolution of $\tau_\mathrm{r}^{-1}$ with exposure time is also shown in figure~\ref{fig:trpl}(b). The most plausible explanation for this decrease in the radiative decay rate of excitons is an increase in the radial electric fields by the C deposition. This effect implies that C adatoms induce surface states that pin the Fermi level closer to midgap compared to the unexposed \emph{M}-plane surfaces. 

The quantities depicted in figure~\ref{fig:trpl}(b) directly determine the internal quantum efficiency $\eta_\mathrm{int} = \tau_\mathrm{eff}/\tau_\mathrm{r}$. In figure~\ref{fig:trpl}(c), we compare $\eta_\mathrm{int}$ deduced from the time-resolved PL experiments to the CL intensity $I_\mathrm{CL}$ recorded during the exposure of the six investigated fields, which is expected to be proportional to $\eta_\mathrm{int}$. Note that $\eta_\mathrm{int}$ is likely to be different for steady-state and transient conditions and is certain to depend on excitation density \cite{brandt_prb_1996a}. Also, the depth profile of the excitation differs between PL and CL. Considering the dissimilarity in experimental conditions, the agreement between these two different measurements is surprisingly good. The faster initial decay in CL can be attributed to charge trapping, while the slower decay for long times may result from the high excitation density in CL experiments.
  
\section{Summary and conclusion}

We have shown that both charge trapping at the nanowire sidewall surfaces and the deposition of a carbonaceous film due to the cracking of hydrocarbons by the electron beam contribute to the luminescence quenching of GaN nanowires under electron beam exposure. The former effect is fast, reversible, and comparatively weak. The latter is persistent, acts on longer time scales, and quenches the CL intensity by up to two orders of magnitude. We point out that these processes may be interlinked, i.\,e., hydrocarbon fragments created in the vacuum may actually be attracted to the scanned area due to surface charging.

We have ruled out the possibility of optical absorption in the deposited carbonaceous film. Instead, this film has been found to induce strong nonradiative surface recombination and, at the same time, to drastically reduce the radiative decay rate of free excitons. These two observations suggest that C adatoms create surface states that are highly efficient nonradiative centers and simultaneously pin the Fermi level closer to mid gap, thus increasing the radial electric fields in the nanowires. To further elucidate the role of C on GaN surfaces, it would be interesting to employ surface sensitive techniques to  clarify the detailed chemical and electronic configuration for C adsorbed on GaN surfaces of different orientations.

The luminescence quenching already occurs for sub-monolayer C coverage and will be difficult to avoid even with the utmost care in sample handling and vacuum conditions, unless an ultra-high vacuum setup is used. As a consequence, the quenching effect has to be accounted for in the investigation of nanowires by CL spectroscopy. In particular, the relative height of different spectral contributions can be significantly altered, because emission from strongly localized states, such as impurity-related transitions, is much less affected than the NBE luminescence arising (at room temperature) from free excitons. Moreover, in time-resolved CL measurements, the transients are distorted by the buildup of C on the surface during the masurement. It is also important to take great care when nanostructures are imaged by SEM prior to subsequent PL measurements. In particular, carrier lifetimes obtained by such measurements may not represent those of the pristine nanostructure, but may rather reflect surface modifications as observed in the present work for GaN nanowires.

Nanostructures of other semiconductors, such as ZnO, are likely  to suffer from similar surface effects during electron beam exposure. For GaN layers, the surface clearly plays a less significant role. Nevertheless, at low acceleration voltages and thus shallow penetration depths, surface effects become important also for CL measurements on layers, and the effects discussed in this work should play a role there as well.

\ack
The authors would like to thank Carsten Pfüller for a critical reading of the manuscript.

\bibliography{quenching}

\end{document}